\documentclass[a4paper,reprint,unsortedaddress,prl]{revtex4-1}

\usepackage[english]{babel}
\usepackage[utf8x]{inputenc}
\usepackage[T1]{fontenc}

\usepackage{amssymb}
\usepackage{amsmath}
\usepackage{amsthm}
\usepackage{graphicx}
\usepackage{tikz}
\usepackage[shortlabels]{enumitem}
\usepackage[normalem]{ulem}
\PassOptionsToPackage{hyphens}{url} 
\usepackage[breaklinks=true]{hyperref}

\theoremstyle{plain}

\theoremstyle{remark}

\theoremstyle{definition}

\renewcommand{\d}{\,\mathrm{d}} 

\newcommand{\eps}{\varepsilon}
\newcommand{\vks}{v_\mathrm{KS}^\eps}
\newcommand{\rhoks}{\rho_\mathrm{KS}^\eps}

\DeclareMathOperator{\trace}{Tr}

\begin{document}
\title{Guaranteed Convergence of a Regularized Kohn--Sham Iteration in Finite Dimensions}

\author{Markus Penz}
\affiliation{Max Planck Institute for the Structure and Dynamics of Matter, Hamburg, Germany}

\author{Andre Laestadius}
\affiliation{Hylleraas Centre for Quantum Molecular Sciences, Department of Chemistry, University of Oslo, Norway}

\author{Erik I. Tellgren}
\affiliation{Hylleraas Centre for Quantum Molecular Sciences, Department of Chemistry, University of Oslo, Norway}

\author{Michael Ruggenthaler}
\affiliation{Max Planck Institute for the Structure and Dynamics of Matter, Hamburg, Germany}

\begin{abstract}
    The exact Kohn--Sham iteration of generalized density-functional theory in finite dimensions with a Moreau--Yosida regularized universal Lieb functional and an adaptive damping step is shown to converge to the correct ground-state density.
\end{abstract}
\maketitle

The Kohn--Sham (KS) scheme~\cite{KS1965} of ground-state density-functional theory (DFT) is the cornerstone of electronic structure calculations in quantum chemistry and solid-state physics \cite{burke2012perspective}. It maps a complicated system of interacting electrons onto an auxiliary, non-interacting KS system. This yields a set of coupled one-particle equations that need to be solved self-consistently. Since a direct solution is unfeasible, practical approaches are variations of self-consistent field methods taking the form of fixed-point iterations or energy minimization algorithms~\cite{KARLSTROM_CPL67_348,RABUCK_JCP110_695,CANCES_MMNA34_749,HELGAKER_CPL327_397,KUDIN_JCP116_8255,HOST_JCP129_124106}. To date, no method has been rigorously shown to converge to the correct ground-state density. Convergence results for approximate schemes are available for auxiliary assumptions~\cite{liu2015analysis}, and reliably achieving convergence in systems with small band gaps or for transition metals remains a hard practical challenge~\cite{KUDIN_ESAIM41_281}.
Approximation techniques face the problem of an exponential growth of local minima with increasing number of particles~\cite{FUKUTOME_PTP45_1382}. Such local minima appear as `false' solutions in the energy landscape and distract from the global, absolute minimum \cite{stillinger1999exponential}.
Hence, a method with mathematically guaranteed convergence to the correct minimizer is of central importance and has been listed as one of twelve outstanding problems in DFT \cite{ruzsinszky2011twelve}.

An early insight is that iterations commonly fail unless oscillations between trial states are damped~\cite{KARLSTROM_CPL67_348}. Work by Canc{\`e}s and Le Bris~\cite{cances2000can,cances2001} led to the optimal damping algorithm (ODA) based on energy minimization by line search along the descent direction. \citet{Wagner2013,wagner2014kohn} presented a similar scheme and claimed to have proven convergence in the setting of exact DFT, while only the strict descent of energies was secured.
In such efforts, functional differentiability is almost always tacitly assumed or wrongly claimed, prominently in Ref.~\cite[Eq.~(2.105)]{engel2011density}, while the underlying universal functionals are known to be non-differentiable \cite{Lammert2007}. This means the usual presentations of DFT already \emph{assume} some form of regularization of the functionals.
Other special forms of DFT like with internal magnetic fields \cite{tellgren2018density} or finite temperatures \cite{chayes1985density,giesbertz2019one} automatically include regularization effects.

This issue was addressed in \citet{KSpaper2018}, where a similar iterative scheme was proposed that proved a weak type of convergence after Moreau--Yosida (MY) regularization to ensure differentiability of the universal Lieb functional~\cite{Lieb1983}.
Weak-type convergence here means that the energy converges to either the correct energy or an upper bound. MY regularization has been introduced to DFT by \citet{Kvaal2014}.

A rich study of possible strategies for self-consistent field iteration was recently put forward by \citet{lammert2018bivariate}. Yet in all those works the question of a limit density and corresponding KS potential was left open. On the other hand, the result in \citet{KSpaper2018} is applicable to not only standard DFT, but to all DFT flavors that fit into the given framework of reflexive Banach spaces. It has already been successfully applied to paramagnetic current DFT  (CDFT)~\cite{MY-CDFTpaper2019}. This general approach is also pursued in this Letter.

In what follows we give a fully rigorous proof of convergence for the KS scheme in a finite-dimensional state space.
Because of the techniques involved, the new iteration scheme was baptized ``MYKSODA'' in \citet{MY-CDFTpaper2019}.
The employed damping critically depends on MY regularization that bounds the curvature of the universal Lieb functional from above.

We will now present the mathematical framework. For a much more detailed discussion of generalized KS schemes in Banach spaces that can also be infinite dimensional, we refer to \citet{KSpaper2018}.
The spaces for densities and potentials are chosen to be the Hilbert space $X=X^*=\ell^2(M)$, $M \in \mathbb{N}$, which corresponds to a finite one-particle basis, a lattice system with $M$ sites, or many other possible settings.
The reason for this dual choice of spaces is how densities and potentials couple in the energy expression. What is denoted $\int_{\mathbb R^3} v\rho \d x$ in standard DFT is a finite sum in the given setting and will further be written $\langle v,\rho \rangle$ with $\rho \in X, v\in X^*$.
For the internal energy of the \emph{full system}, a universal functional $\tilde F$, like the one defined by constrained search~\cite{levy1979,Lieb1983} over all $N$-particle density matrices $\Gamma$ that yield a given density $\rho\in X$, is introduced,
\begin{equation}
\tilde F(\rho) = \inf_{\Gamma \mapsto \rho}\left\{ \trace((H_\mathrm{kin}+H_\mathrm{int})\Gamma) \right\}.
\end{equation}
Here $H_\mathrm{kin}$ stands for the kinetic energy and $H_\mathrm{int}$ for interactions.
Consequently, the functional $\tilde F$ is defined on a set $\tilde X \subset X$ of \emph{physical densities} that come from an $N$-particle density matrix (ensemble $N$-representability).
This set $\tilde X$ will be assumed bounded in $X$. Since all physical densities are normalized in the $\ell^1$ norm and all norms are equivalent in finite dimensions, this follows naturally. It also holds for CDFT on a finite lattice, since the current density is bounded by the hopping parameter~\cite[Eq.~(25)]{farzanehpour2012lattice}, and for 
one-body reduced density matrix functional theory (RDMFT) in finite basis sets, since the off-diagonal elements of the reduced density matrix are bounded by the diagonal ones that give the usual density~\cite[Eq.~(3.49)]{giesbertz2019one}.

On the other hand, elements in $X$ will in general not constitute physical densities. In standard DFT this means that an arbitrary $x \in X$ does not have to be normalized or even positive. Such an $x \in X$ will thus be called a \emph{quasi-density}. We reserve the notation $\rho$ for physical densities.

The total energy is the infimum of $\tilde F(\rho)$ plus the the potential energy coming from a given external potential $v \in X^*$, taken over all physical densities,
\begin{equation}\label{eq:E-def}
E(v) = \inf_{\rho \in \tilde X} \{ \tilde F(\rho) + \langle v,\rho \rangle \}.
\end{equation}
It is linked to a functional $F$ on $X$ by the Legendre--Fenchel transformation (convex conjugate). Then $F$ can be transformed back to the same $E$ as
\begin{align}
&F(x) = \sup_{v\in X^*} \{ E(v) - \langle v,x \rangle \}, \\
&E(v) = \inf_{x \in X} \{  F(x) + \langle v,x \rangle \}. \label{eq:LF-EF}
\end{align}
The functional $F$ is by construction convex and lower semi-continuous~\cite[Th.~3.6]{Lieb1983} and has $F(x)=+\infty$ whenever $x$ is not in the domain $\tilde X$ of $\tilde F$. Minimizers of \eqref{eq:E-def} are the \emph{ground-state densities}, which establishes a link to the Schrödinger equation. They stay the same if one switches from $\tilde F$ to $F$, and thus minimizers of \eqref{eq:LF-EF} are always in $\tilde X$. Finding such minimizers $\rho$ of \eqref{eq:LF-EF} is equivalent to determining the superdifferential of $E$, i.e., the set of functionals in $X^{**}=X$ that yield a graph completely above $E$,
written $\rho \in \overline{\partial}E(v) \subset \tilde X$.

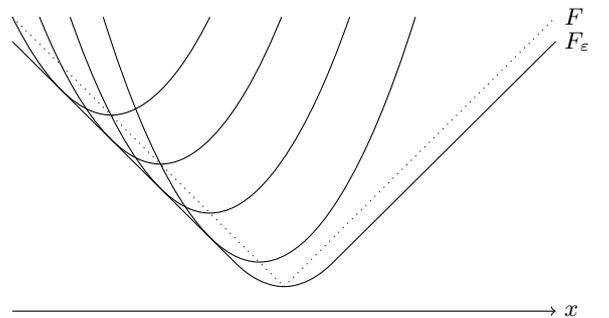
\begin{figure}[ht]
\centering
\begin{tikzpicture}[scale=.65]
    \draw[->] (-1,0) -- (10,0)
        node[right] {$x$};
    
    \draw (-1,6) parabola bend (1,4) (3,6);
    \draw (-.45,6) parabola bend (2,3) (4.45,6);
    \draw (.17,6) parabola bend (3,2) (5.83,6);
    \draw (.84,6) parabola bend (4,1) (7.16,6);
    
    \draw (-1,5.5) -- (3.5,1);
    \draw (3.5,1) parabola bend (4.5,.5) (5.5,1);
    \draw (5.5,1) -- (10,5.5) node[right] {$F_\eps$};
    
    \draw[dotted] (-1,6) -- (4.5,.5);
    \draw[dotted] (4.5,.5) -- (10,6) node[right] {$F$};
\end{tikzpicture}
\caption{Moreau--Yosida regularization $F_\eps$ of an exemplary $F$, showing also the regularization parabolas $\frac{1}{2\eps}\|x-y\|^2$ that trace out $F_\eps$.}
\label{fig:MY}
\end{figure}

The MY regularization of the functional $F$ on $X$ is defined as
\begin{equation}
F_\eps(x) = \inf_{y\in X} \left\{ F(y) + \tfrac{1}{2\eps}\|x-y\|^2 \right\}.
\end{equation}
The visual understanding of this is the following. As the vertex of the regularization parabola $\frac{1}{2\eps}\|x\|^2$ moves along the graph of $F$, the regularized $F_\eps$ is given by the traced out lower envelope (the ``Moreau envelope'' \cite[Def.~1.22]{rockafellar2009variational}). This is visualized in Fig.~\ref{fig:MY}. It also means the regularization puts an upper bound of $\eps^{-1}$ on the (positive) curvature of $F_\eps$. This will be an important ingredient in the convergence proof: A bound on the curvature means the convex functional $F_\eps$ cannot change from falling to rising too quickly, yielding a secure bound on the possible step length for descent.

The regularized $F_\eps$ is then differentiable and even has a continuous gradient $\nabla F_\eps$ (Fréchet differentiability) \cite[Th.~9]{KSpaper2018}, something that will also become important in the convergence proof. We define the associated energy functional
\begin{equation}\label{eq:def-Eeps}
    E_\eps(v) = \inf_{x \in X} \{  F_\eps(x) + \langle v,x \rangle \}.
\end{equation}
The functional in \eqref{eq:def-Eeps} is \emph{not} the MY regularization of $E$ but the Legendre--Fenchel transformation of $F_\eps$. If $z\in X$ is a minimizer in \eqref{eq:def-Eeps}, called the \emph{ground-state quasi-density}, then the gradient of $F_\eps+v$ at $z$ must be zero,
\begin{equation}\label{eq:diff-Feps}
 \nabla F_\eps(z) + v = 0.
\end{equation}
Since the regularized functional is differentiable everywhere, the usual problem of $v$-representability is avoided.
For $E_\eps$, which is still not differentiable, we can resort to the superdifferential. Since any such element $z \in \overline{\partial}E_\eps(v)$ automatically solves \eqref{eq:diff-Feps}, it is the ground-state quasi-density of the regularized problem with potential $v$. Two important properties of $E_\eps$ are \cite[Th.~10 and Cor.~11]{KSpaper2018}
\begin{align}
    E(v) &= E_\eps(v) + \frac{\eps}{2}\|v\|^2, \label{eq:E-reg-unreg}\\
    \overline{\partial}E(v) &= \overline{\partial}E_\eps(v) + \eps v \subset \tilde X, \label{eq:dE-reg-unreg}
\end{align}
which relate the regularized problem back to the unregularized one. Since $E$ is already concave, the subtraction of a parabola in \eqref{eq:E-reg-unreg} makes $E_\eps$ \emph{strongly} concave. Note that the $\eps$ in \eqref{eq:dE-reg-unreg} takes a role comparable to that of \emph{permittivity}, linking potentials to densities.

To set up a KS scheme we define a \emph{reference system} that is non-interacting by
\begin{equation}
\tilde F^0(\rho) = \inf_{\Gamma \mapsto \rho}\left\{ \trace(H_\mathrm{kin}\Gamma) \right\}
\end{equation}
on the same $\tilde X \subset X$ and define $E^0, F_\eps^0,$ and $E_\eps^0$ analogously. The analogue of \eqref{eq:diff-Feps} for $F_\eps^0$ at the same quasi-density $z\in X$ is
$\nabla F^0_\eps(z) + \vks = 0$
and defines the KS potential $\vks$.
Simply equating this equation and \eqref{eq:diff-Feps} gives
\begin{equation}\label{eq:diff-F-eq}
 \nabla F_\eps(z) + v = \nabla F^0_\eps(z) + \vks,
\end{equation}
where the ground-state quasi-density $z$ and the auxiliary $\vks$ for the reference system are still unknown and neither $F_\eps$ nor $F^0_\eps$ have a simple, explicit expression.
The trick is to determine $z$ and $\vks$ in an iterative algorithm by replacing them with sequences $x_i \to z$, $v_i \to \vks$. The indicated convergence is our major concern in the following proof. We get an update rule for the potential sequence $(v_i)_i$ directly from \eqref{eq:diff-F-eq},
\begin{equation}\label{eq:eq:step-a}
 v_{i+1} = v + \nabla F_\eps(x_i) - \nabla F^0_\eps(x_i),
\end{equation}
and determine the next quasi-density by solving for the ground-state of the regularized reference system with $v_{i+1}$.
This iteration has the stopping condition $v_{i+1} = -\nabla F_\eps^0(x_i)$, which means $v = -\nabla F_\eps(x_i)$ by \eqref{eq:eq:step-a}. Then $x_i$ is already the sought-after ground-state quasi-density $z$ and thus also $v_{i+1}=\vks$ is the respective KS potential that yields the same quasi-density for the reference system.

The most important ingredient of \emph{practical} KS calculations enters by giving suitable approximations for the expression $\nabla F_\eps - \nabla F^0_\eps$ (Hartree-exchange-correlation (Hxc) potential including the correlated kinetic energy). For the purpose of showing convergence it is not crucial that this object comes from the exact functional or that it is the result of an approximation, as long as $F_\eps,F^0_\eps$ have the stated properties.

The MYKSODA algorithm is then the following. In step (a) get the new potential by \eqref{eq:eq:step-a} above.
In step (b) solve the (simpler) ground-state problem for the reference system by choosing the next quasi-density from $\overline\partial E_\eps^0(v_{i+1})$.
From \eqref{eq:dE-reg-unreg} it follows that the set of quasi-densities $\overline\partial E_\eps^0(v_{i+1})$ can be determined from the set of ground-state densities of the reference system, which means solving the non-interacting Schrödinger equation.
Finally, to ensure a strictly descending energy and to show convergence of $(x_i)_i,(v_i)_i$, we include a damping step (c) with an adaptively chosen step length.

\emph{MYKSODA iteration scheme.}---Assume $\tilde X$ bounded and $E^0$ finite everywhere.
For $v\in X^*$ fixed, set $v_1=v$ and select $x_1\in \overline\partial E_\eps^0(v)$. Iterate $i=1,2,\dots$ according to

\begin{enumerate}[(a)]
	\item set $v_{i+1} = v + \nabla F_\eps(x_i) - \nabla F_\eps^0(x_i)$ and stop if $v_{i+1} = -\nabla F_\eps^0(x_i) = \vks$,

	\item select $ x_{i+1}' \in\overline\partial E_\eps^0(v_{i+1})$ and get the step direction $y_i = (x_{i+1}'-x_i)/\|x_{i+1}'-x_i\|$,

	\item choose the step length
	$\tau_i = -\eps \langle \nabla F_\eps(x_i) + v, y_i \rangle > 0$ and set $x_{i+1} = x_i + \tau_i y_i$.
\end{enumerate}

We prove below that this algorithm guarantees convergence to the correct KS potential, $v_i \to \vks$, and to the ground-state quasi-density, $x_i \to z$, of both the full system with $v$ and the reference system with $\vks$.
The corresponding energy is then determined by $E_\eps(v) = F_\eps(z)+\langle v,z \rangle > -\infty$.
These are still solutions of the regularized problem, but with \eqref{eq:E-reg-unreg} and \eqref{eq:dE-reg-unreg} a transformation back to the unregularized setting is easily achieved. This, unlike the usually assumed unregularized KS scheme, gives \emph{different} ground-state densities for the non-interacting and the interacting system, while circumventing all problems of differentiability and thus of $v$-representability.
The assumption that $E^0$ is finite everywhere is trivially fulfilled in a finite-dimensional setting because $E^0$ is a finite sum. It is still kept here to connect more closely to standard DFT and CDFT, where $E^0$ is finite even in the infinite-dimensional setting, see \citet[Th.~3.1(iii)]{Lieb1983} and \citet[Lem.~20]{MY-CDFTpaper2019}, respectively.

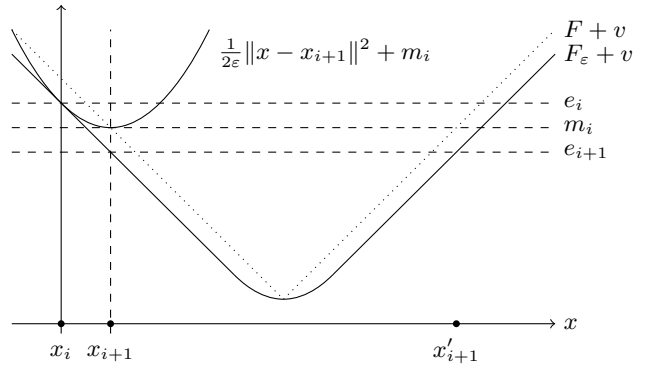
\begin{figure}[ht]
\centering
\begin{tikzpicture}[scale=.65]
    \draw[->] (-1,0) -- (10,0)
        node[right] {$x$};
    \draw[->]
        (0,-0.2) -- (0,6.5);
    
    \fill (0,0) circle [radius=2pt] node[below=5pt] {$x_i$}; 
    \fill (1,0) circle [radius=2pt] node[below=5pt] {$x_{i+1}$};
    \fill (8,0) circle [radius=2pt] node[below=2pt] {$x'_{i+1}$};
    \draw[dashed] (1,-.2) -- (1,6);
    
    \draw (-1,6) parabola bend (1,4) (3,6) node[below right] {$\tfrac{1}{2\eps} \|x-x_{i+1}\|^2 + m_i$};
    
    \draw (-1,5.5) -- (3.5,1);
    \draw (3.5,1) parabola bend (4.5,.5) (5.5,1);
    \draw (5.5,1) -- (10,5.5) node[right] {$F_\eps+v$};
    
    \draw[dotted] (-1,6) -- (4.5,.5);
    \draw[dotted] (4.5,.5) -- (10,6) node[right] {$F+v$};
    
    \draw[dashed] (-1,4.5) -- (10,4.5) node[right] {$e_i$};
    \draw[dashed] (-1,4) -- (10,4) node[right] {$m_i$};
    \draw[dashed] (-1,3.5) -- (10,3.5) node[right] {$e_{i+1}$};
\end{tikzpicture}
\caption{Illustration of one iteration step.}
\label{fig:MYKSODA}
\end{figure}

\emph{Convergence proof.}---We refer to the the first part of the proof of Theorem~12 in \citet{KSpaper2018} to show that the superdifferential $\overline\partial E_\eps^0(v_{i+1})$ is everywhere non-empty because of $E^0$ finite, guaranteeing that the (regularized) ground-state problem in (b) always has at least one solution.
The directional derivative of $F_\eps+v$ at $x_i$ in direction $x'_{i+1} - x_i$ can be rewritten by (a),
\begin{equation}
\label{eq:new-eq-ref}
\langle \nabla F_\eps(x_{i}) +v, x'_{i+1} - x_i \rangle 
    = \langle v_{i+1} + \nabla F_\eps^0(x_i), x'_{i+1} - x_i \rangle.
\end{equation}
Realizing that $x'_{i+1} \in \overline{\partial}E_\eps^0(v_{i+1})$ from (b) and $x_i \in \overline{\partial}E_\eps^0(\nabla F_\eps^0(x_i))$ from invertibility \cite[Lem.~4]{KSpaper2018}, we rewrite the right-hand side of \eqref{eq:new-eq-ref} with the help of \eqref{eq:dE-reg-unreg}, substituting
\begin{align}
    &\tilde x'_{i+1} = x'_{i+1} + \eps v_{i+1} \in \overline{\partial}E^0(v_{i+1}), \\
    &\tilde x_i = x_i + \eps \nabla F_\eps^0(x_i) \in \overline{\partial}E^0(\nabla F_\eps^0(x_i)),
\end{align}
which gives
\begin{equation}
\langle v_{i+1} + \nabla F_\eps^0(x_i), \tilde x'_{i+1} - \tilde x_i \rangle - \eps\|v_{i+1} + \nabla F_\eps^0(x_i)\|^2.
\end{equation}
Now, since $\tilde x'_{i+1}, \tilde x_i$ are selected from the superdifferential of $E^0$ for the respective potentials $v_{i+1}, \nabla F_\eps^0(x_i)$, the inner product is always smaller or equal to zero \cite[Lem.~5]{KSpaper2018}. This property is called monotonicity of $\overline{\partial}E^0$ and directly follows from concavity of $E^0$. What follows is \emph{strong} monotonicity of $\overline{\partial}E^0_\eps$, i.e.,
\begin{equation}\label{eq:strong-mon-hs}
\begin{aligned}
    \langle \nabla F_\eps(x_{i}) +v, x'_{i+1} - x_i \rangle 
    &= \langle v_{i+1} + \nabla F_\eps^0(x_i), x'_{i+1} - x_i \rangle \\
    &\leq  - \eps\|v_{i+1} + \nabla F_\eps^0(x_i)\|^2 \\
    &= - \eps\|\nabla F_\eps(x_{i}) +v\|^2.
\end{aligned}
\end{equation}
The last line follows from (a) and is strictly smaller than zero if not $\|\nabla F_\eps(x_{i}) +v\| = 0$, which would mean that we have already converged to the ground-state quasi-density. We thus infer that, unless converged, we always have a negative directional derivative of $F_\eps+v$ at $x_i$ in the step direction $y_i$ which is parallel to $x'_{i+1} - x_i$, i.e.,
$\langle \nabla F_\eps(x_{i}) +v, y_i \rangle < 0.$
Such a negative directional derivative means the left leg of the regularization parabola is aligned tangentially to the (differentiable) energy functional $F_\eps + v$, like depicted in Fig.~\ref{fig:MYKSODA}. The next quasi-density $x_{i+1} = x_i + \tau_i y_i$ is then chosen at the vertex of this regularization parabola.
%
%
This corresponds to a choice of step length $\tau_i$ where the directional derivatives at $x_i$ in direction $y_i$ of the regularization parabola $\frac{1}{2\eps}\|\cdot - x_{i+1}\|^2$ and of $F_\eps + v$ are equal,
\begin{equation}\label{eq:derivative-tau-hs}
\begin{aligned}
\langle \nabla F_\eps(x_i) + v, y_i \rangle &= -\frac{1}{\eps} \langle x_{i+1}-x_i, y_i \rangle \\ &= -\frac{1}{\eps} \|x_{i+1}-x_i\| = -\frac{\tau_i}{\eps}.
\end{aligned}
\end{equation}
This construction yields a $x_{i+1} \neq x_i$, where the energy $e_i = F_\eps(x_i) + \langle v,x_i \rangle$ is always larger than the energy value $m_i$ at the vertex, see Fig.~\ref{fig:MYKSODA}.
Since the regularization parabola lies fully above the energy functional $F_\eps+v$ by construction, the energy $e_{i+1}$ at $x_{i+1}$ must obey $e_{i+1} \leq m_i < e_i$. The strictly decreasing $e_i$ is now by definition bounded below by $E_\eps(v)$ from \eqref{eq:def-Eeps} and thus converges.
By determining $e_i-m_i$ from the regularization parabola and then combining it with \eqref{eq:derivative-tau-hs},
\begin{equation}
  \frac{\tau_i^2}{2\eps} = \frac{1}{2\eps} \|x_{i+1}-x_i\|^2 = e_i-m_i    \leq e_i-e_{i+1}\to 0,
\end{equation}
we can infer convergence of $(x_i)_i$. Step (a) then defines an associated potential 
\begin{equation}
\lim_{i\to\infty} v_{i+1} = v + \nabla F_\varepsilon^1(x) - \nabla F_\varepsilon^0(x),
\end{equation}
since the gradients are both continuous.
After having proved that the densities and potentials converge, it shall be demonstrated that they converge to the expected ground-state quasi-density $z$ and KS potential $\vks$.
We come back to \eqref{eq:strong-mon-hs}, where substituting $x_{i+1}'-x_i = y_i \, \|x_{i+1}'-x_i\|$ gives
\begin{equation}
    \|x_{i+1}'-x_i\| \, \langle \nabla F_\eps(x_{i}) +v, y_i \rangle 
    \leq -\eps \|\nabla F_\eps(x_{i}) +v\|^2,
\end{equation}
which together with \eqref{eq:derivative-tau-hs} results in
\begin{equation}
\|x_{i+1}'-x_i\| \, \frac{\tau_i}{\eps} \geq \eps \|\nabla F_\eps(x_{i}) +v\|^2.
\end{equation}
We already know from the convergence of densities that $(x_i)_i$ is bounded, further
\begin{equation}
x'_{i+1} \in \overline\partial E_\eps^0(v_{i+1}) = \overline\partial E^0(v_{i+1}) - \eps v_{i+1},
\end{equation}
by (b) and \eqref{eq:dE-reg-unreg}. But $\overline\partial E^0(v_{i+1}) \subset \tilde X$, which is bounded, and $(v_i)_i$ converges as well.
Thus, $\|x_{i+1}'-x_i\|$ is bounded, and since $\tau_i \to 0$, it follows $\|\nabla F_\eps(x_{i}) +v\| \to 0$ and $\|\nabla F_\eps^0(x_i) + v_{i+1}\| \to 0$. This in turn means $v = -\nabla F_\eps(\lim x_i)$, so $\lim x_i=z$ is the ground-state quasi-density for the potential $v$ in the full, regularized problem. Finally, $\lim v_{i+1} = -\lim \nabla F_\eps^0(x_i) = -\nabla F_\eps^0(z) = \vks$ is the KS potential.
$\Box$

As noted above, the reference system reproduces the quasi-density $z$ of the full system and they link back to the real densities by \eqref{eq:dE-reg-unreg},
\begin{equation}\label{eq:dens-reg-unreg}
    \rho = z + \eps v, \quad
    \rhoks = z + \eps \vks,
\end{equation}
where typically $\rhoks \neq \rho$. Then $\vks-v=\eps^{-1}(\rhoks-\rho)$ is precisely the Hxc potential that depends on the regularization parameter $\eps$ here. This means every choice of $\eps$ defines a \emph{different} reference system. A limit $\eps \to 0$ in the algorithm is unfeasible because of its relation to the step length.

\begin{figure}
\includegraphics[width=1\linewidth]{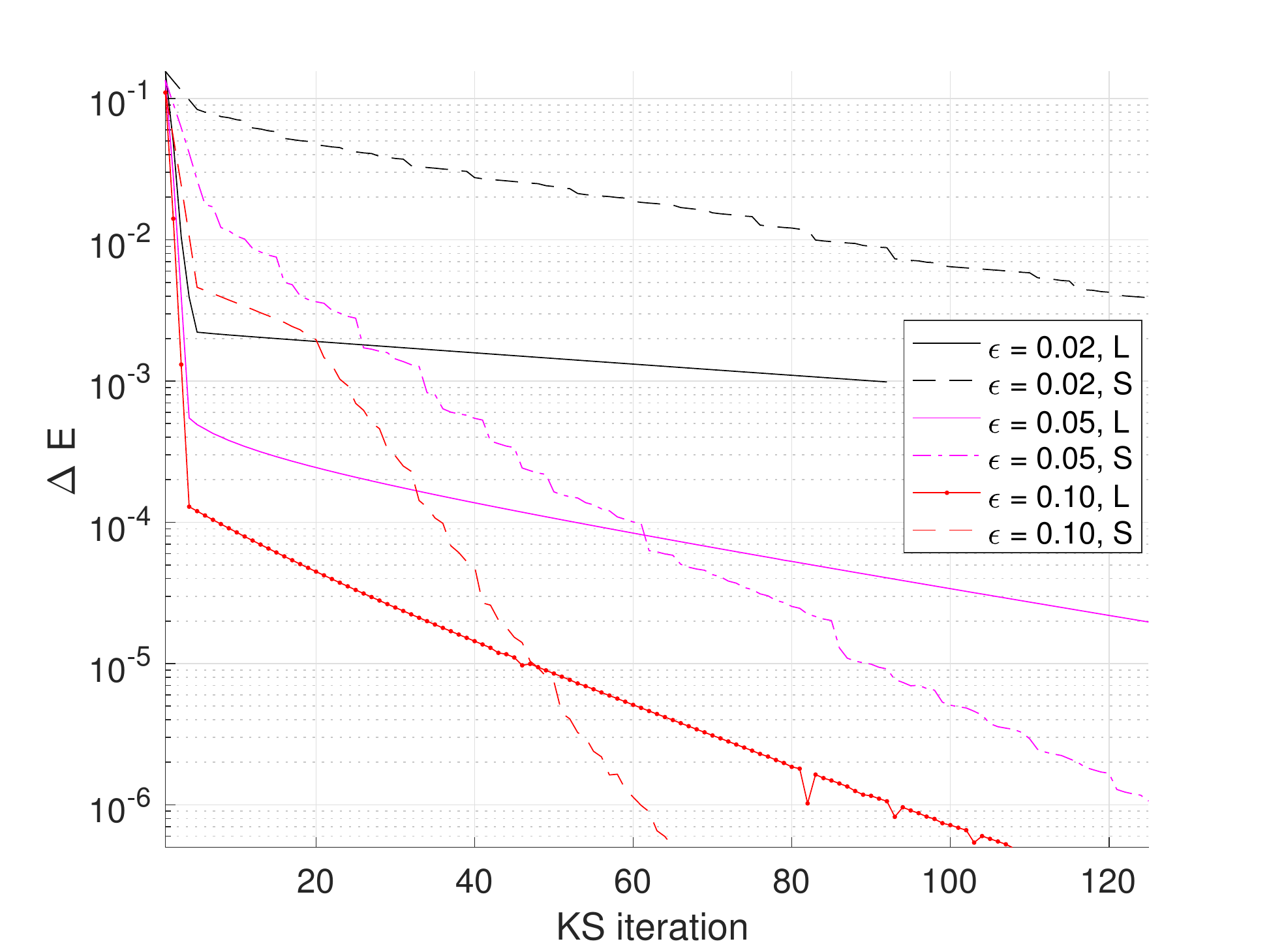}
\caption{Convergence of $\Delta E_i = e_i - E_\eps(v)$ for a ring-lattice system with external potential $v=\cos(2\theta)+0.2\cos(\theta)$ and different $\eps$. The algorithm developed here is labeled ``S'' while the method ``L'' chooses the step length maximally \cite{KSpaper2018,MY-CDFTpaper2019}.}
\label{fig:energy-conv}
\end{figure}

A simulation of two electrons on a ring lattice~\cite{MYringProgram} allows us to illustrate the above method. Compared to a previous implementation in a CDFT setting~\cite{MY-CDFTpaper2019}, the version given here uses the more conservative damping step that helped to prove convergence. To distinguish the two versions, we denote them ``MYKSODA-S'' for shorter, conservative steps, and ``-L'' for the original longer steps~\cite{KSpaper2018,MY-CDFTpaper2019}. Both versions have been adapted to a pure DFT setting, $H_\mathrm{kin}$, taking the form of a standard second-order finite difference. A radius of $R=1$~bohr, a uniform grid with 30 points, and the interaction energy $H_{\mathrm{int}}=3 \sqrt{1+\cos(\theta_1-\theta_2)}$ were used. As expected, larger $\varepsilon$ leads to faster convergence. Also, the more conservative steps taken by MYKSODA-S often lead to slower convergence in practice. Surprisingly, however, in some cases MYKSODA-S overtakes the less conservative MYKSODA-L. An example is shown in Fig.~\ref{fig:energy-conv}. Such a crossover is possible as the two algorithms follow different paths through the space of densities and potentials. Yet, when the starting point is the same, the \emph{first} step by MYKSODA-L always lowers the energy more than the first step by MYKSODA-S. Although it is a plausible conjecture that also MYKSODA-L, taking maximal steps, is guaranteed to converge, the present proof does not establish this.
``Maximal steps'' here means taking $\tau_i$ maximally such that $\langle \nabla F_\eps(x_{i+1})+v,x'_{i+1}-x_i \rangle \leq 0$, which yields maximal decrease in energy in the direction chosen by step (b).

In this Letter we proved convergence of the regularized KS scheme with special adaptive damping. In short, this means that KS-DFT is a veritable method to calculate the correct ground-state density. This strong statement holds for all flavors of DFT that are defined on a finite-dimensional density space $X=\ell^2(M)$ and have a linear coupling to external potentials of type $\langle v,\rho \rangle$. This includes CDFT, where the potential $v$ is a combination of scalar and vector potential, and the density $\rho$ includes the paramagnetic current density. To allow for a combination of these different entities into one Banach space setting, the respective function spaces for one-particle densities and current densities have to fulfill a condition termed ``compatibility'' in \citet{MY-CDFTpaper2019}. A proof of MYKSODA convergence for infinite-dimensional Banach spaces $X$ is feasible but much more technical and will be presented elsewhere. The choice of step length in (c) is an essential part of the proof and similar choices could be of value in showing convergence of related iteration schemes and for other settings such as Hartree--Fock theory~\cite{cances2000can}. Next to this damping step the MY regularization is a vital part of the proof at hand, not only to have functional differentiability, but also for the strong monotonicity estimate needed to show convergence. How those findings can be transferred to realistic KS implementations will be the content of future research, but it is expected that they serve as useful guidelines for better convergence results.

\begin{acknowledgments}
\emph{Acknowledgments.}
We express our gratitude for fruitful discussion with Simen Kvaal. 
MP is grateful for the hospitality received at the Hylleraas Centre for Quantum Molecular Sciences in Oslo and furthermore acknowledges support by the Erwin Schr\"odinger Fellowship J 4107-N27 of the FWF (Austrian Science Fund).
ET and AL were supported by CoE Hylleraas Centre for Molecular Sciences Grant No.~262695, as well as the Norwegian Supercomputing Program (NOTUR) through a grant of computer time (Grant No.~NN4654K).
ET acknowledges support from Research Council of Norway through Grant No.~240674. AL acknowledges support from European Research Council under Grant Agreement No.~ERC-STG-2014 639508.
\end{acknowledgments}

%

\newpage

\section*{Erratum et Corrigendum}

\vspace{-.5em}
\noindent
\textbf{Additional author:}
Paul E. Lammert,
\emph{Department of Physics, Pennsylvania State University, USA}\\

The central argument for convergence of the ``MYKSODA iteration scheme'' introduced in our article \cite{KS_PRL_2019} contains a critical mistake that we hereby want to correct. The claim was that because of $\frac{1}{2\eps}\|x_{i+1}-x_i\|^2 \leq e_i-e_{i+1}\to 0$ in Eq.~(19) the sequence of quasidensities $(x_i)_i$ also converges. This is wrong, since this does not show the necessary Cauchy property $\|x_m-x_n\| \to 0$ for all $m,n \to \infty$. Convergence could be secured, on the other hand, in cases where $e_i - e_{i+1} \leq r^i$ with $0<r<1$, because then $\|x_{i+1}-x_i\| \leq C \sqrt{r}^i$ and summability of the geometric series would give the Cauchy property for $(x_i)_i$. Yet, such a condition for the convergent energy sequence $(e_i)_i$ is in general not available. The corrigendum given below proceeds along a different path that effectively takes degeneracies of the ground state into account. First, it is shown that the sequence of quasidensities $(x_i)_i$ is bounded. Since we operate in a finite-dimensional setting this means by the Bolzano--Weierstrass theorem that the sequence has at least one convergent subsequence. We then show that all such subsequences converge to ground-state quasidensities of the regularized, fully interacting problem. Thereby, the accumulation points of the sequence give the desired solutions of which there can be more than one in case of degeneracy. Although in this way the algorithm can accommodate multiple valid solutions, we cannot be sure that we map out \emph{all} possible solutions. The selection of specific solutions might still depend on the initial value $x_0$.\\

While the MYKSODA iteration scheme remains precisely the same, the statement following it has to be modified. We prove that for any convergent subsequence $x_{\alpha(i)} \to z$ the corresponding potential sequence $v_{\alpha(i)}$ converges to the correct KS potential that reproduces $z$ in the regularized reference problem.
We will now repeat the last part of the proof which needs to be adjusted.\\

\emph{Corrected convergence proof.}
Up to Eq.~(19) all statements stay intact, just the inference of convergence right after Eq.~(19) is erroneous.
The $F_\eps$ increases like $\|x\|^2$ asymptotically by construction (cf.\ the regularization procedure in Eq.~(5)) because $F(x)=+\infty$ whenever $x$ is outside the bounded $\tilde X$. Since $e_i=F_\eps(x_i)+\langle v,x_i \rangle \leq e_1$ for all $i$ this means that $(x_i)_i$ must already be bounded. This translates to boundedness of $(v_i)_i$ defined by step (a), because $\nabla F_\eps, \nabla F^0_\eps$ are Lipschitz-bounded (with constant $\eps^{-1}$; \citet[Corollary 2.59]{Barbu-Precupanu}). We can then argue for $\|\nabla F_\eps(x_i)+v\| \to 0$ and $\|\nabla F^0_\eps(x_i) + v_{i+1}\| \to 0$, exactly like in the original proof. This means $v = -\lim \nabla F_\eps(x_i)$ and for any accumulation point $z$ of $(x_i)_i$ (Bolzano--Weierstrass guarantees that there is at least one) there must be a convergent subsequence $x_{\alpha(i)} \to z$ such that $v = -\lim \nabla F_\eps(x_{\alpha(i)}) = -\nabla F_\eps(z)$. By inversion we get $z \in \overline\partial E_\eps(v)$ which means that $z$ is the correct ground-state quasidensity of the regularized full problem. Finally, $\lim v_{\alpha(i+1)} = -\lim \nabla F^0_\eps(x_{\alpha(i)}) = -\nabla F^0_\eps(z) = \vks$ is the KS potential that exactly reproduces this $z$ in the regularized reference problem. A different accumulation point with its associated convergent subsequence $(x_{\beta(i)})_i$ will lead to another possible ground-state quasidensity of the regularized full problem in the case of degeneracy. Then the procedure assigns it the appropriate KS potential as the limit of $(v_{\beta(i)})_i$ that reproduces the same quasidensity in the regularized reference problem. In this manner the whole sequence $(x_i)_i$ partitions into convergent subsequences that have ground-state quasidensities as limit points.  $\Box$\\

Note that it is still true that the distance between successive terms goes to zero, i.e.\ $\|x_{i+1}-x_i\| \to 0$. This means that no isolated accumulation points are possible. But since the ground-state solutions in terms of density matrices $\Gamma$ form a convex set and the mapping $\Gamma \mapsto \rho$ is linear, also the set of ground-state (quasi)densities will be convex and thus corresponds to a whole `accumulation region'. The sequence $(x_i)_i$ will thus converge to this set of solutions. If the ground-state density is non-degenerate (unique) then $\overline\partial E_\eps(v)$ is single valued and the whole sequence converges to this point only.\\

Another small mistake should be noted: After (13), two times in (15), and after (16) there must be a minus sign in front of $\nabla F^0_\eps$. The error does not influence any subsequent results. We would also like to point out a misleading feature of Fig.~1, where the impression is given that the minima of the one-dimensional sections through $F+v$ and $F_\eps+v$ coincide, which holds true in the special case of the absolute minima but not along an arbitrary direction.
   
\vspace{-1.5em}

\end{document}